\documentclass{aa}  

\usepackage{graphicx}
\usepackage{txfonts}
\begin{document}

   \title{Mind the trap}
   \subtitle{Non-negligible effect of volatile trapping in ice on C/O ratios in protoplanetary disks and exoplanetary atmospheres}

   \author{N.F.W. Ligterink
          \inst{1,2}
          \and
          K.A. Kipfer \inst{1}
          \and 
          S. Gavino \inst{3}
          }

 \institute{
    Space Research \& Planetary Sciences, Physics Institute, University of Bern, Sidlerstrasse 5, CH-3012 Bern Switzerland 
    \and
    Faculty of Aerospace Engineering, Delft University of Technology, Delft, The Netherlands \\
             \email{niels.ligterink@tudelft.nl}
    \and
        Niels Bohr Institute, University of Copenhagen, Øster Voldgade 5-7, 1350, Copenhagen K, Denmark
}

   \date{Received April 16, 2024; accepted June 17, 2024}

% \abstract{}{}{}{}{} 
% 5 {} token are mandatory
 
  \abstract
   {}
   {The ability of bulk ices (H$_{2}$O, CO$_{2}$) to trap volatiles has been well studied in any experimental sense, but largely ignored in protoplanetary disk and planet formation models as well as the interpretation of their observations. We demonstrate the influence of volatile trapping on C/O ratios in planet-forming environments.}
   {We created a simple model of CO, CO$_{2}$, and H$_{2}$O snowlines in protoplanetary disks and calculated the C/O ratio at different radii and temperatures. We included a trapping factor, which partially inhibits the release of volatiles (CO, CO$_{2}$) at their snowline and  releases them instead, together with the bulk ice species (H$_{2}$O, CO$_{2}$) . Our aim has been to assess its influence of trapping solid-state and gas phase C/O ratios throughout planet-forming environments.}
   {Volatile trapping significantly affects C/O ratios in protoplanetary disks. Variations in the ratio are reduced and become more homogeneous throughout the disk when compared to models that do not include volatile trapping. Trapping reduces the proportion of volatiles in the gas and, as such, reduces the available carbon- and oxygen-bearing molecules for gaseous accretion to planetary atmospheres. Volatile trapping is expected to also affect the elemental hydrogen and nitrogen budgets.}
   {Volatile trapping is an overlooked, but important effect to consider when assessing the C/O ratios in protoplanetary disks and exoplanet atmospheres. Due to volatile trapping, exoplanets with stellar C/O have the possibility to be formed within the CO and CO$_{2}$ snowline.}


   \maketitle
%
%-------------------------------------------------------------------

\section{Introduction}

\begin{figure}[ht]
\centering
\includegraphics[width=\hsize]{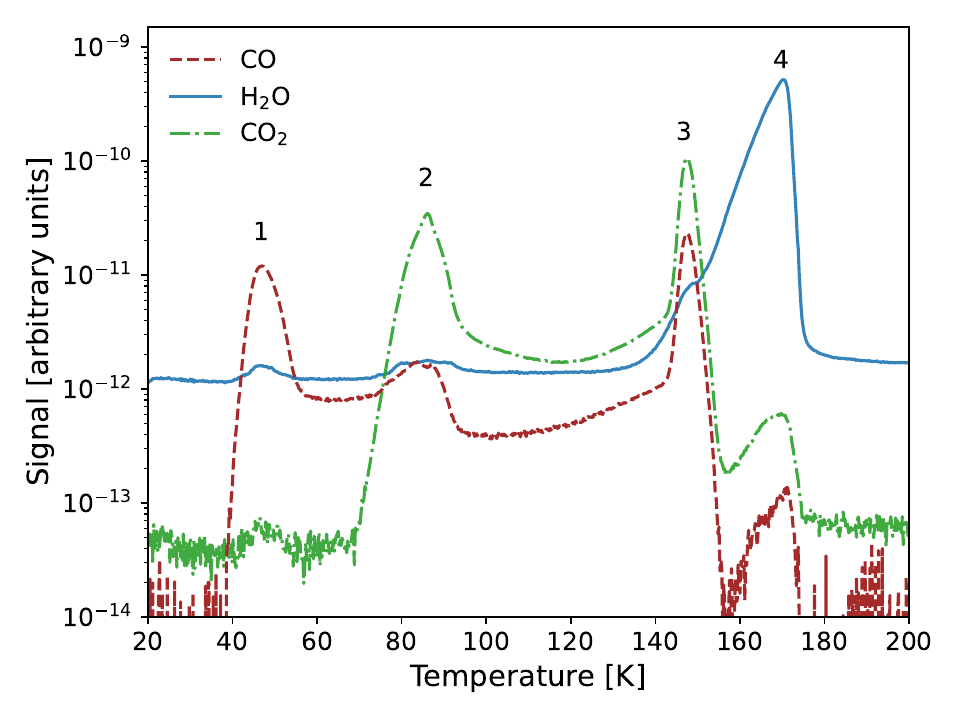}
  \caption{Thermal desorption of a mixed H$_{2}$O:CO$_{2}$:CO ice film at 100:15:5 ratio recorded during a Temperature Programmed Desorption (TPD) experiment, where the release of each molecule is measured with a mass spectrometer. Carbon monoxide was traced with the $^{13}$CO isotope. Four distinct release events are visible: pure-phase CO (1), pure-phase CO$_{2}$ with CO co-desorbing (2), volcano desorption of CO and CO$_{2}$ (3), and water desorption with co-desorption of CO and CO$_{2}$ (4). Desorption and trapping fractions of volatiles can be determined by integrating the TPD trace over discrete temperature ranges. The data are adapted from \citet{kipfer2024sublimation}.} 
  \label{fig:tpd}
\end{figure}

Water (H$_{2}$O), carbon dioxide (CO$_{2}$), and carbon monoxide (CO) are some of the most abundant molecules found in the interstellar medium and planet-forming environments, specifically on ice-coated dust grains \citep{boogert2015}. Due to their prominence and widespread availability, the physicochemical processes involving these molecules have been extensively studied in the laboratory \citep[e.g.][]{hama2013surface,vandishoeck2013interstellar,linnartz2015}. This includes their thermal desorption behaviour and the interplay of these molecules during desorption. Their desorption temperatures differ significantly, with the monolayer coverage peak desorption temperatures measured in the laboratory, ranging from $\sim$30--40~K for the hypervolatile CO, $\sim$80--90~K for CO$_{2}$, and to around 155--175~K for H$_{2}$O \citep{minissale2022}. In fact, out of all prominent ice mantle components (e.g. also including CH$_{4}$, NH$_{3}$, and CH$_{3}$OH), water is the least volatile. Combined with its high abundance, H$_{2}$O ice serves as an important surface onto which molecules can adsorb and react, but also as a medium into which volatile species can be trapped \citep{Barnun_1985,collings2004}. This way of locking up atoms and molecules is sometimes listed as mechanical trapping, entrapment, or physical trapping, but it is always related to the ability of a bulk medium to cover up a single or cluster of species and preserve them in the ice matrix. Laboratory experiments have shown that large fractions of volatile molecules can be trapped well above their peak desorption temperature. An example is given in Fig. \ref{fig:tpd}, which shows a temperature programmed desorption (TPD) trace of a mixed H$_{2}$O:CO$_{2}$:CO ice adapted from \citet{kipfer2024sublimation}. As the ice is linearly heated, the release of these molecules is followed with a mass spectrometer. The molecules sublimate at their characteristic pure temperatures (1, 2, 4), but CO and CO$_{2}$ also release well above their desorption temperature. Carbon monoxide co-releases with CO$_{2}$ (2), while CO and CO$_{2}$ both also release with the water sublimation event. This happens in the form of co-desorption with H$_{2}$O at $\sim$170~K (4) and in the form of a volcano desorption event at $\sim$145~K (3), which occurs when water-ice rearranges from an amorphous to crystalline structure and volatile species are expulsed from the ice \citep{burke2010}. Water ice can trap fractions of over tens of percent and up to a hundred percent of added volatiles, depending on the molecule or atom involved, ice structure, ice formation temperature, mixing ratios, and deposition rate \citep{Barnun_1985,notesco2003gas,collings2004,fayolle2011laboratory,Almayrac_2022,Simon_2023,kipfer2024sublimation}. Similar trapping efficiencies have been found for CO$_{2}$ ice as bulk medium \citep{simon2019entrapment}.  A more detailed description of how desorption and trapping fractions are determined in laboratory experiments is provided in Appendix \ref{ap:trapping}. The trapping of volatiles affects the physical and chemical processes in star and planet-forming environments, but nuanced differences are expected between trapping in a laboratory and natural environment. For example, \citet{ciesla2018efficiency} showed that efficient trapping takes place when the timescale for a volatile species to desorb is longer than the time it takes to be covered by a monolayer of water; this is the so-called `burial~ regime'. These conditions are met at high deposition rates, as usually used in laboratory experiments, or when surface temperatures are low. Considering the low temperatures at which H$_{2}$O formation takes place on interstellar dust grains \citep{hama2013surface}, it is expected that experimentally determined trapping efficiencies apply to interstellar environments. However, slow deposition in interstellar environments and at elevated temperatures will likely result in different trapping efficiencies.

Planets are thought to form in protoplanetary disks by the accretion of gas and ice-coated dust grains, pebbles, or planetesimals \citep[e.g.][]{Johansen+2007, Lambrechts+2012, Bitsch+2019, Lambrechts+2019}. The bulk molecular species contained in solids and in gas play a prominent role in setting the elemental composition of a planet. \citet{oberg2011effects} presents a model of the radially varying gas phase and solid-state carbon-over-oxygen (C/O) ratio in a protoplanetary disk, due the step-wise release of CO, CO$_{2}$, and H$_{2}$O from the ice to the gas. The model predicts that the C/O ratio of a planetary atmosphere depends on which side of the major volatile (CO/CO$_{2}$/H$_2$O) snowlines the object is formed. A high C/O ratio ($>$ 0.8) indicates that the formation was initiated outside the snowlines prior to inward migration \citep{Madhusudhan+2014}. An exoplanet transit is typically used to derive the atmospheric composition and a significant number of C/O ratio values have now been measured \citep{Hoch+2023}. The \citet{oberg2011effects} model has been used to determine the formation location of exoplanets based on their observed atmospheric C/O ratio \citep{molliere2020retrieving,zhang202113co}. For instance, the supersolar gas phase C/O ratio ($\gtrapprox$ 0.55) outside the water snowline was used to explain the C/O $\geq$ 1 ratio measured in the atmosphere of the exoplanet WASP-12b \citep{madhusudhan2011high}. Conversely, the low C/O ratio measured in the atmosphere of WASP-77Ab is interpreted as a formation scenario where the atmospheric material is accreted inside the major snowlines \citep{Line+2021}. In the last decade, more refined models to predict protoplanetary disk snowlines and C/O ratios have been presented in the literature, which includes solid-state and gas phase chemical processes, grain growth, or novel physical processes \citep{owen2020snow,notsu2020composition,cridland2020impact,zhang2020excess,eistrup2022chemical,gavino2023shaping}. However, the trapping of volatiles within the matrix of bulk H$_{2}$O and/or CO$_{2}$ has largely been ignored in the models \citep{schneeberger2023evolution}, despite that formalisms for volatile trapping have been included in a select number of astrochemical models \citep{viti2004evaporation,visser2009ices,taquet2012multilayer,garrod2022}. 

In this paper, we revisit the static model of \citet{oberg2011effects} and include volatile trapping to show that this has a non-negligible effect on the C/O ratio in a protoplanetary disk and (exo)planets that form there. In Section \ref{sec:model}, we describe the modified model and show how trapping affects solid-state and gas phase C/O ratios in protoplanetary disks. In Section \ref{sec:disc}, these results and their implications for protoplanetary disk and exoplanet atmosphere C/O ratios are discussed. 

%--------------------------------------------------------------------
\section{Model and results}
\label{sec:model}

\begin{figure}
\centering
\includegraphics[width=\hsize]{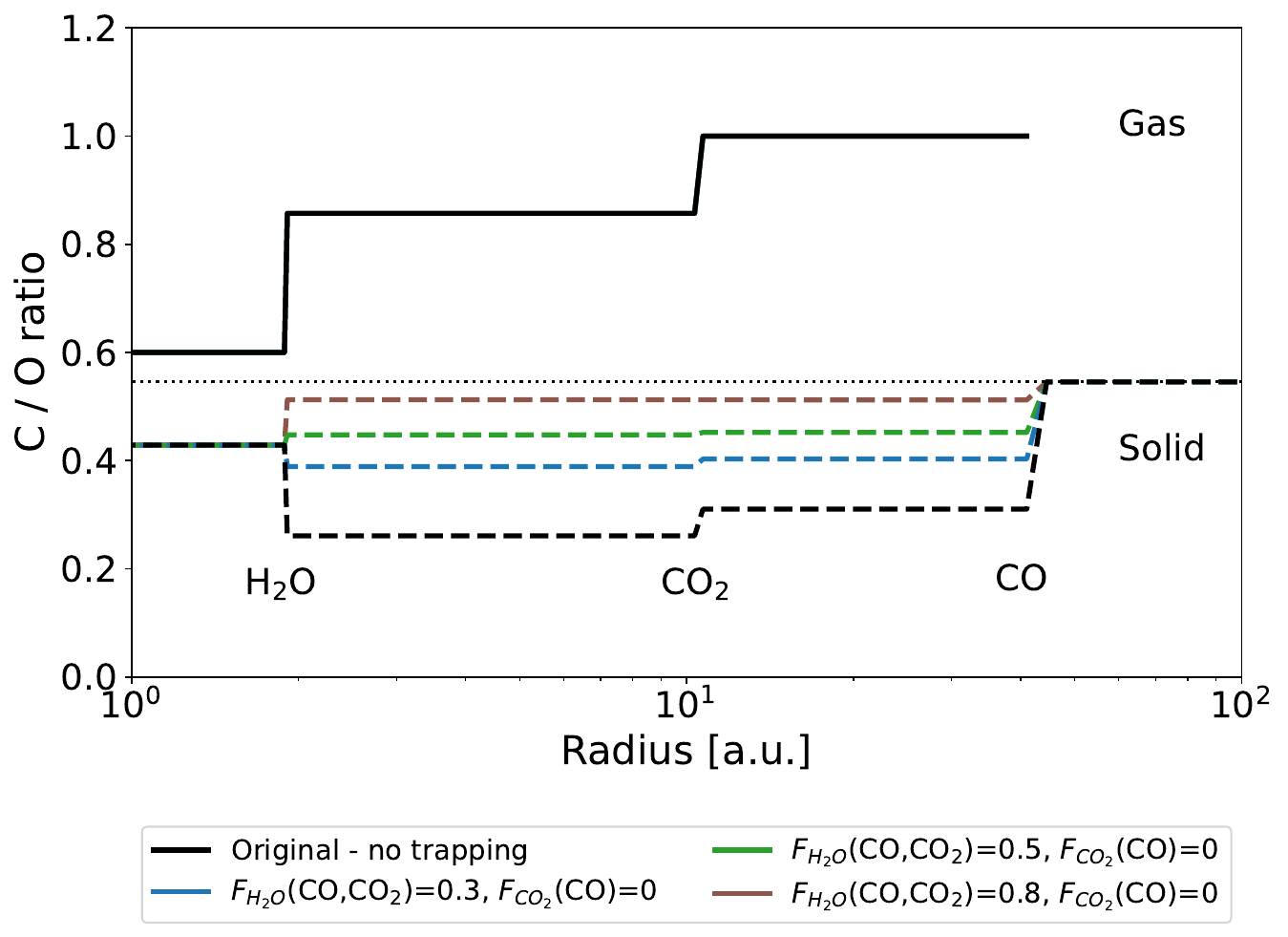}
  \caption{Gas phase and solid-state C/O ratios at various radii in a protoplanetary disk, based on the trapping and release of CO, CO$_{2}$, and H$_{2}$O. The original model of \citet{oberg2011effects} without trapping is presented in black. The updated model takes the trapping of CO and CO$_{2}$ in H$_{2}$O ($F_{\rm H_{2}O}$(CO,CO$_{2}$)) and CO in CO$_{2}$ ($F_{CO_{2}}$(CO)) at different fractions into account (blue, green, and red lines) and shows that the solid C/O ratio is strongly affected. The snowlines of CO, CO$_{2}$, and H$_{2}$O are indicated.
          }
     \label{fig:1}
\end{figure}

To determine the protoplanetary disk C/O ratio, we adopted the static snowline model presented in \citet{oberg2011effects}. In short, this model assumes that CO, CO$_{2}$, and H$_{2}$O ices fully sublimate at distinct sublimation temperatures, that is, their respective snowlines. At each snowline, one molecular species desorbs, which results in the depletion of solid carbon and/or oxygen, while simultaneously setting or modifying the gas phase elemental budget. Carbon grains and silicates are significant elemental carriers, but sublimate at much higher temperatures than H$_{2}$O. Therefore, they remain solid in the temperature range used in this model and only affect the solid carbon and oxygen budget. The model parameters, namely, the sublimation temperatures and elemental abundances per molecule, are presented in Table \ref{tab:model}. The relative abundances between H$_{2}$O, CO$_{2}$, and CO are (in that order) 0.9:0.3:1.5 and are derived from the CBRR 2422.8-3423 protoplanetary disk \citep{pontoppidan2006spatial}. We note that sublimation temperatures differ from the aforementioned laboratory desorption temperatures, due to the difference in heating timescale. Snowlines are plotted on a protoplanetary disk temperature profile following the equation $T = T_{\rm 0} r^{-q}$, where $T_{\rm 0}$ = 200~K, $r$ is the radius in a.u., and $q$ is the power law index of 0.62 \citep{andrews2007}, where smaller radii correspond to higher temperatures and larger radii to lower temperatures. The resulting radial gas phase and solid-state C/O distributions of the original \citet{oberg2011effects} model (black line) are shown in Fig. \ref{fig:1}. This model does not take volatile trapping into account.

\begin{table}[ht]
\centering      
\caption{\label{tab:model} Model parameters}
\begin{tabular}{c c c c}        
\hline\hline                 
Molecule & $T$ & $n_{\rm O}$ & $n_{\rm C}$ \\
& (K) & 10$^{-4}$ $\times$ $n_{\rm H}$ & 10$^{-4}$ $\times$ $n_{\rm H}$ \\
\hline    
CO & 20 & 1.5 & 1.5 \\
CO$_{2}$ & 47 & 0.6 & 0.3 \\
H$_{2}$O & 135 & 0.9 & --  \\
Carbon grains & 500 & -- & 0.6 \\
Silicate & 1500 & 1.4 & -- \\
\hline                                   %inserts single line
\end{tabular}
\tablefoot{Model parameters are adopted from \citet{oberg2011effects}, which in turn are based on data from  \citet{pontoppidan2006spatial}, \citet{draine2003interstellar}, and \citet{whittet2010oxygen}.}
\end{table}

\begin{figure}[ht]
\centering
\includegraphics[width=\hsize]{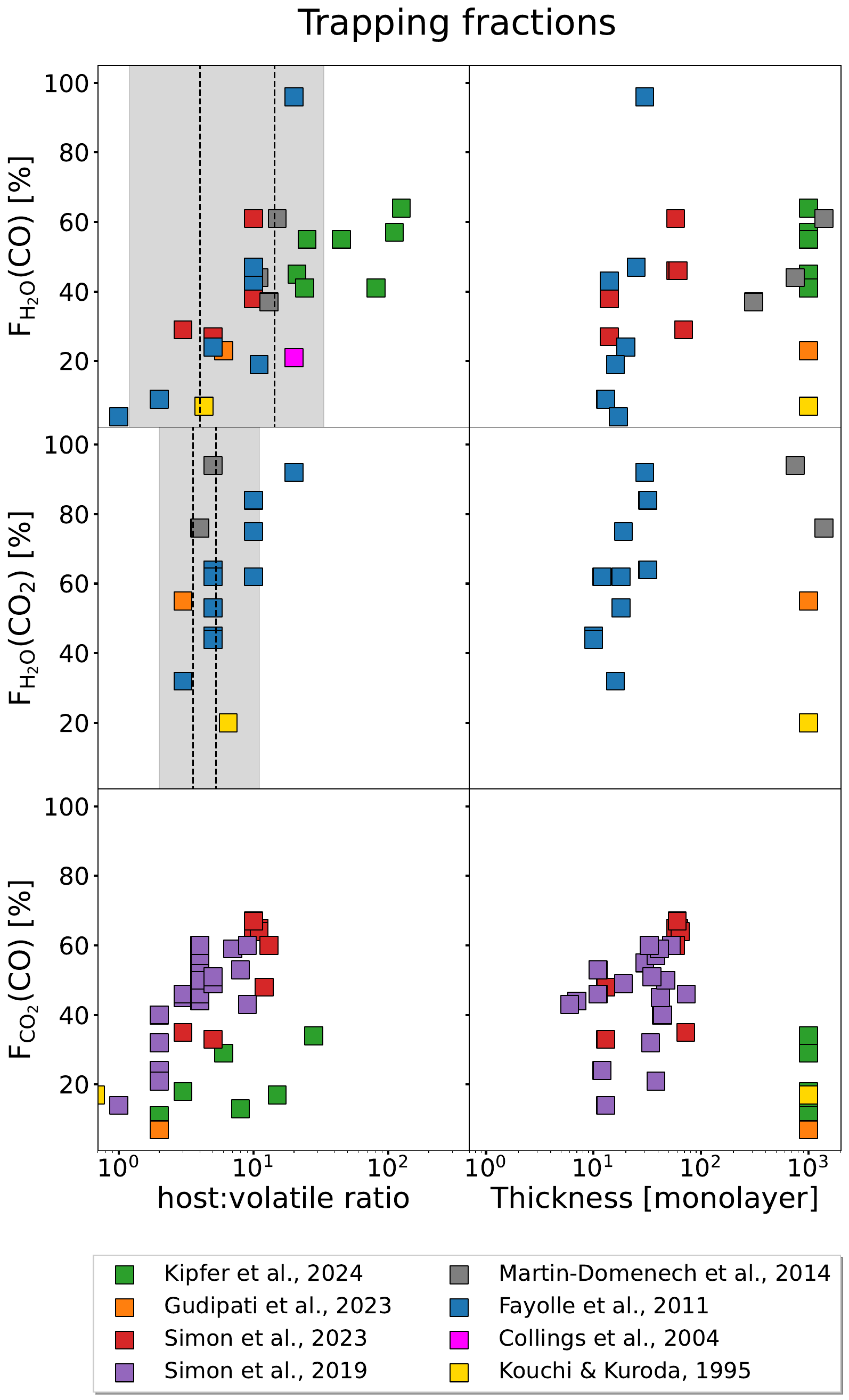}
  \caption{Trapping efficiencies of CO and CO$_{2}$ in water ($F_{\rm H_{2}O}$(CO,CO$_{2}$)), and CO in CO$_{2}$ ($F_{\rm CO_{2}}$(CO)) plotted against the host: volatile ice ratio (left) and the ice layer thickness (right). The thickness is given in monolayer (ML), where 1 ML = 10$^{15}$~molecules~cm$^{-2}$. The grey shaded boxes show the H$_{2}$O:CO and H$_{2}$O:CO$_{2}$ ice ratios observed in a variety of interstellar sources, while the dashed lines show their largest and smallest median values for data presented in \citet{Boogert_2015}. The laboratory data used in this plot are presented in Table \ref{tab:frac}.}
     \label{fig:trap}
\end{figure}

We modified the above model to include volatile trapping by releasing a fraction of the volatiles with water ($F_{\rm H_{2}O}$(CO,CO$_{2}$)) or CO$_{2}$ ($F_{\rm CO_{2}}$(CO)). Volatile release associated with water sublimation occurs in two ways: volcano desorption, namely, the rapid release of volatiles during the amorphous to crystalline phase change of ice, and co-desorption, namely, the release of volatiles together water desorption \citep{burke2010}. Volcano desorption occurs at a lower temperature than the peak water sublimation temperature by approximately $\sim$20\% \citep{collings2004,kipfer2024sublimation}. Experimental studies show that significant volatile release starts with (or is even dominated by) volcano desorption and continues until all volatiles are depleted during the water co-desorption event \citep{Martin_Domenech_2014}. In the modified static model we release all volatiles interior of the water snowline, but we note that under realistic conditions loss of some CO and CO$_{2}$ will also occur leading up to the water snowline. Co-desorption of volatiles with CO$_{2}$ occurs at the CO$_{2}$ sublimation temperature \citep{simon2019entrapment,kipfer2024sublimation} and this is adopted as such in the modified model. 

Over recent decades, many studies have determined which fractions of volatiles are trapped by H$_{2}$O and CO$_{2}$ \citep[e.g.][]{Barnun_1985,collings2004,fayolle2011laboratory,ligterink2018c,Simon_2023,kipfer2024sublimation}. Figure \ref{fig:trap} shows a compilation of the literature results and presents trapping fractions plotted against the matrix: volatile mixing ratio and the ice layer thickness given in monolayers (ML, 1 ML = 10$^{15}$~molecules~cm$^{-2}$). The data and methods to determine trapping fractions are described in Appendix \ref{ap:trapping}. In general, thicker ice films trap more volatiles, while a large volatile concentration results in a smaller fraction of volatiles being trapped. From this overview, we can say that trapping fractions in an ice mantle are $F_{\rm CO_{2}}$(CO) = 0.1 -- 0.7, $F_{\rm H_{2}O}$(CO$_{2}$) = 0.3 -- 0.95, and $F_{\rm H_{2}O}$(CO) = 0.2 -- 0.7 for mixing ratios $\geq$5:1. We test the effect of various realistic trapping fractions in a series of model runs.

Figure \ref{fig:1} shows three models with water trapping fractions of $F_{\rm H_{2}O}$(CO,CO$_{2}$) = 0.3, 0.5, and 0.8. Both CO and CO$_{2}$ are trapped in the same fraction. Trapping by CO$_{2}$ is ignored. We find that trapping has a significant effect on the solid-state C/O ratio, while it does not affect its gas phase counterpart. As larger fractions of CO and CO$_{2}$ are trapped in water ice, the variations in solid-state C/O ratio are reduced and can even be considered constant from outside the CO snowline up to the H$_{2}$O snowline for the largest fraction of trapping. 

\begin{figure}
\centering
\includegraphics[width=\hsize]{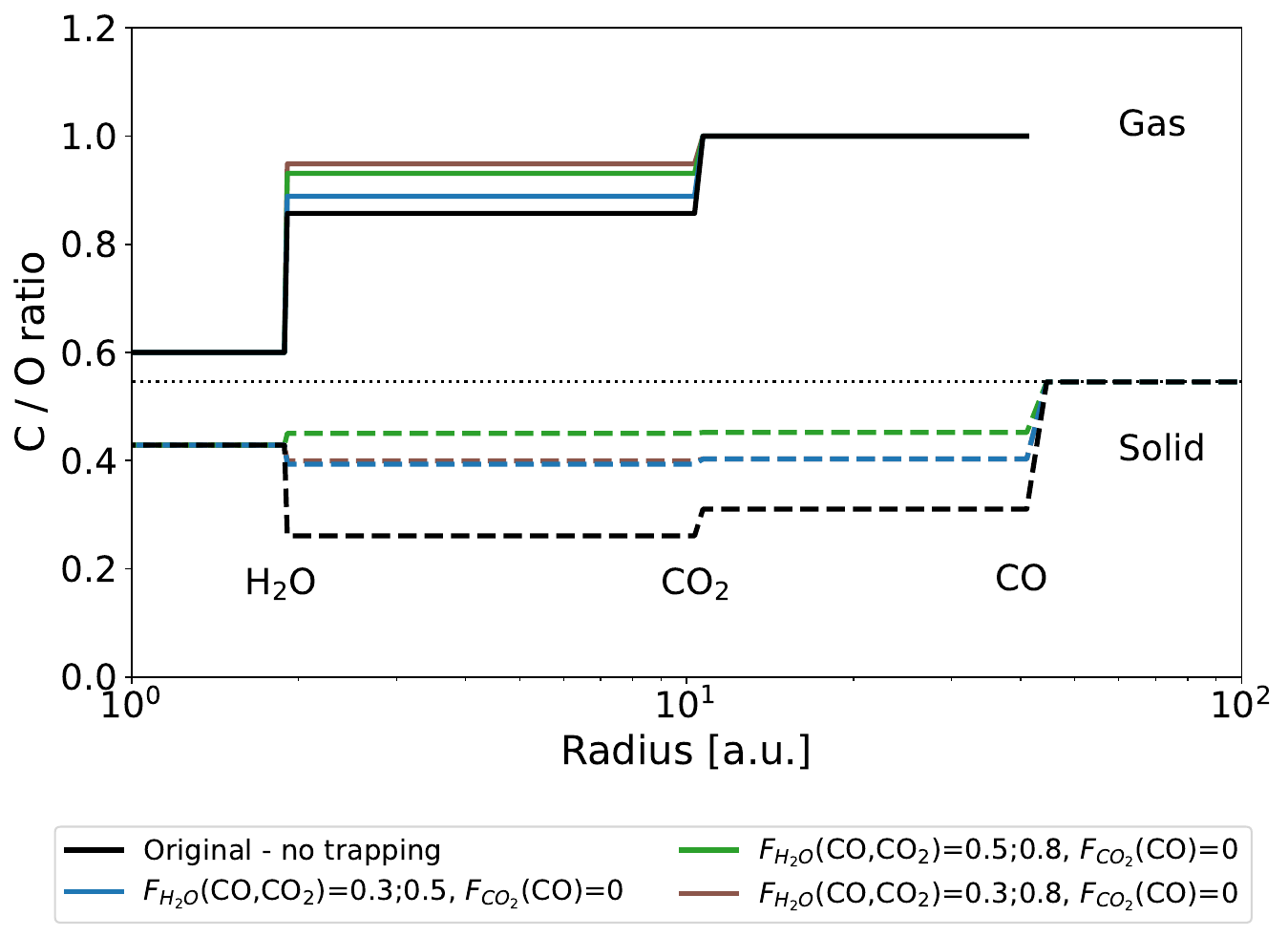}
  \caption{Same as Figure \ref{fig:1}, but different trapping ratios of CO and CO$_{2}$ in H$_{2}$O.
          }
     \label{fig:2}
\end{figure}

Figure \ref{fig:2} shows three models where CO and CO$_{2}$ are trapped in different fractions in water, at $F_{\rm H_{2}O}$(CO,CO$_{2}$) = (0.3;0.5), (0.5;0.8), and (0.3;0.8). In the solid-state, this again results in a reduction in C/O variations. However, this time also the gas phase C/O ratio is affected, which becomes more homogeneous.

Trapping of CO in CO$_{2}$ is included in Figs. \ref{fig:3} and \ref{fig:4}. In general, reduction of variations and homogenisation of the solid-state and gas phase C/O ratios is seen again. The release of a fraction of CO with CO$_{2}$ causes a more pronounced drop in solid-state C/O ratio around the CO$_{2}$ snowline. We conclude that volatile trapping has a non-negligible effect on both solid-state and gas phase C/O ratios.

%--------------------------------------------------------------------
\section{Discussion and Implications}
\label{sec:disc}

In the following sections, we discuss the results and implications from the perspective of interstellar ice compositions and laboratory data on trapping efficiencies, protoplanetary disks, and exoplanet atmospheres. 

%--------------------------------------------------------------------
\subsection{Ice composition and trapping efficiency}
\label{sec:ice}

\begin{figure}[ht]
\centering
\includegraphics[width=\hsize]{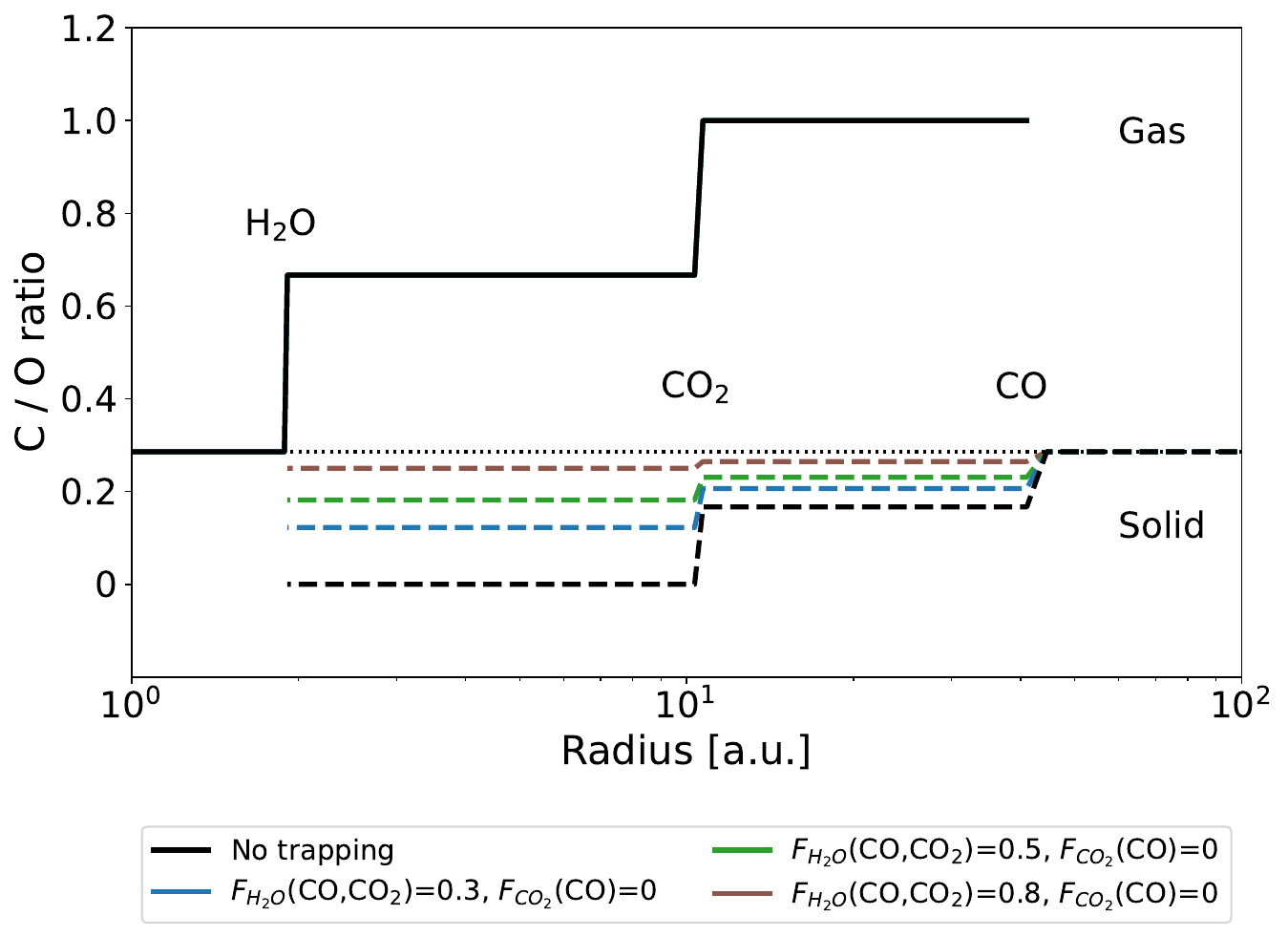}
  \caption{Same as Figure \ref{fig:1}, but only for H$_{2}$O, CO$_{2}$, and CO at 100:25:25 ratio. The contributions of silicates to atomic oxygen and carbonaceous dust to atomic carbon are omitted. 
          }
     \label{fig:5}
\end{figure}

The modified model demonstrates that the trapping of volatiles affects the C/O ratio by reducing variations to this ratio. The extent of this reduction with respect to the solid material depends on the fraction of volatile that is trapped, while for the gas phase, it depends on the size of the difference between trapped fractions of volatiles. Laboratory experiments show that volatile trapping is efficient (see Fig.  \ref{fig:trap}), but this efficiency decreases as the volatile concentration increases. For example, \citet{fayolle2011laboratory} measured a CO trapping fraction in water of $F_{\rm H_{2}O}$(CO) = 0.04 for a 1:1 H$_{2}$O:CO ice film of 17 ML. Since the ice composition used in \citet{oberg2011effects} is 0.9:0.3:1.5 H$_{2}$O:CO$_{2}$:CO, that is, CO is $\sim$1.7 times more abundant than H$_{2}$O, this led to the claim that volatile trapping is negligible. While this might be the case for this particular ice composition, this mixing ratio is not representative of the ice generally found in the interstellar medium or protoplanetary disks \citep{boogert2015}. For example, the median ice compositions presented in \citet{oberg2011spitzer} range from 100:38:31 H$_{2}$O:CO$_{2}$:CO for cloud cores to 100:13:13 for high-mass protostars. Similar patterns are seen for protoplanetary disks \citep{aikawa2012akari,sturm2023jwst}, where H$_{2}$O is the dominant ice component. Figure \ref{fig:5}, shows the model results for a realistic ice composition of H$_{2}$O:CO$_{2}$:CO at a 100:25:25 ratio, both with and without trapping. The contributions of oxygen from silicates and carbon from carbonaceous dust are omitted in this model. For mixing ratios of 4:1 H$_{2}$O to CO$_{2}$ or CO,  $F_{\rm H_{2}O}$(CO$_{2}$) $\geq$40\% and $F_{\rm H_{2}O}$(CO) $\geq$25\% have bee reported \citep{fayolle2011laboratory,Simon_2023}. As can be seen in Fig. \ref{fig:5}, a trapping efficiency of 30\% significantly affects the C/O ratio and causes the same kind of homogenisation as observed in the fiducial model. 

Trapping of volatiles depends on the structure of the ice. For example, when volatile species are covered by a layer of water instead of mixed, the volatile can be fully trapped \citep[][where trapping is defined as the release of volatiles during volcano and co-desorption with water sublimation]{may2013bottom,may2013top}. Astrochemical models have demonstrated that volatiles can be concentrated in the bottom layers of the ice mantle and subsequently be covered by H$_{2}$O as an ice mantle grows \citet{garrod2022}. Furthermore, a recent experimental work by \citet{potapov2023formation} has shown that the UV irradiation of carbonaceous dust covered by H$_{2}$O ice results in significant formation of CO$_{2}$ at the dust-ice interface. Thus, covering of volatiles on ice-coated dust grains is plausible in planet-forming regions and results in trapping fractions that are substantially larger than those of mixed ice. 

Most experimental trapping studies make use of binary ice mixtures and show that volatile trapping is efficient. However, this efficiency (that is, that of the trapped fraction) can decrease as more volatiles are included in the ice. For example, \citet{kipfer2024sublimation} shows that the fraction of N$_{2}$ trapped in water-ice is influenced by the amount of CO$_{2}$ that is included in N$_{2}$:CO$_{2}$:H$_{2}$O mixtures and decreases as more CO$_{2}$ is incorporated. This observation has been linked to competition for binding sites when multiple volatile species are present in a water-ice matrix \citep[e.g.][]{Simon_2023,ligterink2024sublimation}. Ice mantles of grains in planet-forming environments often consist of more than five dominant components and understanding the trapping behaviour of such multicomponent mixtures is relevant to assessing how trapping affects C/O ratios. However, the number of studies that have been conducted on ice systems that approach the complexity of interstellar ice mantles is limited. For example, analysis of TPD data of H$_{2}$O:CO:CO$_{2}$:CH$_{3}$OH:NH$_{3}$ mixed ice systems presented in \citet{Martin_Domenech_2014} suggests that trapping fractions of CO and CO$_{2}$ are similar to those of binary and ternary systems (see Fig. \ref{fig:trap} and Appendix \ref{ap:trapping}). However, dedicated investigations of multi-component ice films are needed to determine if this observation holds and how trapping efficiencies are affected by various ice components and compositions.

In this study, we focus on the elemental carbon and oxygen budget derived from H$_{2}$O, CO$_{2}$, and CO. However, molecules such as methane (CH$_{4}$) should be considered as well. This volatile molecule can also be efficiently trapped in H$_{2}$O and CO$_{2}$ \citep{Simon_2023}. Furthermore, elemental ratios involving atomic nitrogen and hydrogen are also of interest and carriers of these atoms have also been shown to show efficiently in terms of trapping \citep[e.g. N$_{2}$, see][]{kipfer2024sublimation}. 

%--------------------------------------------------------------------
\subsection{Protoplanetary disk snowlines and missing volatiles}
\label{sec:ppd}

Volatile trapping potentially has two prominent observable effects on protoplanetary disks. First, it reduces the proportion of volatiles, such as CO, which are available in the gas phase in regions where the volatile should not be frozen out. Simultaneous measurements of HD (hydrogen deuteride) and CO in disks indicate that these objects are depleted in CO by a factor 5 -- 100 \citep{mcclure2016mass, zhang2020rapid}.  To account for this depletion, various effects have been proposed, such as pebble growth, physical sequestration, and chemical processes \citep[e.g.][]{krijt2018transport,eistrup2018molecular}. Models that combine these processes can predict a CO depletion factor close to 100 \citep{Krijt+2020}. Volatile trapping in ice will contribute to such depletion with trapping fractions of up to 95\% (see Fig. \ref{fig:trap}) for thin film ice. However, orders of magnitude depletion are unlikely for thin ice films and are likely to require thicker ice mantles as found on pebbles \citep{bosman2018efficiency}.

Second, trapping tends to `inject' volatiles at multiple locations in the disk into the gas phase, resulting in a step-wise increase of volatile abundances towards the protostar. For example, CO will enter the gas phase when its pure-phase ice sublimates, when CO$_{2}$ desorbs, and at the onset of H$_{2}$O sublimation. Observations of such step-wise increases might be challenging, due to limited spatial resolution, optically thick emission, excitation temperatures, and the usage of indirect chemical tracers \citep[e.g. N$_{2}$H$^{+}$,][]{van2017robustness}. However, hints of it might be seen in some studies, such as observations performed by \citet{zhang2020excess}, where an enhanced gas phase CO level and C/H ratio were found inside the CO snowline of the HD 163296 protoplanetary disk. These authors assigned this observation to pebble drift, but an alternative explanation relates to the release of carbon-bearing volatiles during the water ice volcano desorption event (see Fig. \ref{fig:tpd}). This increases the gaseous carbon budget, while gas phase oxygen and hydrogen levels remain comparatively low because H$_{2}$O does not yet sublimate.

Other mechanisms can contribute to a step-wise increasing emission or double emission line pattern in a disk. The presence of pure multiple snowlines of one volatile species can be due to the presence of multiple dust surface temperatures. Using a two-sized model, \citet{gavino2023shaping} predicted two pure CO snowlines in a T Tauri disk, the first one at a few tens of au, the second at 100-150 au from the central star. This effect does not require radial drift and relies only on the presence of multiple dust temperatures. The presence of multiple dust sizes and temperatures can also contribute to create a snowline shape that is spread out radially and vertically \citep{Gavino+2021} rather than marking an abrupt step-wise emission. Radial drift can also affect volatile snowlines. \citet{Cleeves_2016}, using a simple parameterised model, showed that the removal of large grains in the outer disk reshapes the thermal structure that can create multiple CO snowlines.

Trapped volatiles may be chemically processed into more complex molecules and thus removed from the ice. As ice-coated grains near a protostar, their temperatures and amounts of impinging radiation increase, which results in enhanced chemical reactions \citep{potapov2023formation}. Under these conditions, CO embedded in H$_{2}$O-ice is efficiently converted into CO$_{2}$ \citep{terwisschavanscheltinga2022}. In turn, CO$_{2}$ thermally reacts with NH$_{3}$ to form carbamic acid \citep[NH$_{2}$COOH,][]{khanna1999carbamic,noble2014,marks2023thermal}. The formation of much more complex and refractory molecules is possible as well \citep[e.g.][]{qasim2023,Kipfer_2023}, which will desorb at much higher temperatures ($\sim$200 -- 400~K) and much closer to protostar \citep{ligterink2023overview}.

Finally, it is worth pointing out that recent edge-on observations of HH 48 NE protoplanetary disk with JWST revealed that CO ice features extend to large disk heights, where the temperature is likely to be high enough that pure, surface-bound, CO ice should have sublimated \citep{sturm2023jwst}. One interpretation of this observation is that CO is trapped in other bulk ice components. This notion is supported by the fact that the CO ice IR feature observed in the disk is broader than that of pure CO measured in the laboratory, which can be the result of interactions of CO with a H$_{2}$O-rich environment. Therefore, this observation highlights the importance of volatile trapping and the effects it may have on the disk chemistry. 

%--------------------------------------------------------------------
\subsection{C/O ratio in exoplanet atmospheres}
\label{sec:exoplanet}

Exoplanets are thought to inherit their elemental abundances from the gas and solids at the location where they form in the parent protoplanetary disk. As such, the measured relative abundance of their chemical content can be used as a diagnostic of their formation pathway. In particular, the \citet{oberg2011effects} model shows that the C/O ratio in hot Jupiters atmospheres is expected to be a signpost for their formation location relative to the main volatile snowlines in the disk. The planets that accrete their envelope inside the water snowline result in an oxygen-rich atmosphere (assuming a carbon-depleted disk) and a subsolar C/O ratio  \citep[e.g.][]{Cridland+2016, Mordasini+2016}, which favours in situ formation scenarios. The model also predicts that atmospheres with superstellar C/O ratios imply the accretion of gas between the H$_{2}$O and CO snowline (and subsequent inward migration after the gas is dissipated), substellar C/O ratios imply the accretion of solids between the H$_{2}$O and CO snowline, and a stellar C/O ratio suggests that the exoplanet formed outside the CO snowline \citep[e.g.][]{madhusudhan2011high,Madhusudhan+2014,Madhusudhan+2017,molliere2020retrieving,zhang202113co, Line+2021}. However, when volatile trapping is included, we find that solid C/O ratios outside and inside the CO snowline are similar, see Fig. \ref{fig:1}. Therefore, the observation of a stellar C/O ratio in an exoplanet atmosphere does not automatically imply that this object was formed outside the CO snowline. Instead, stellar C/O can likely be acquired due to the formation of the planet between the H$_{2}$O and CO snowline. Furthermore, this can affect the ice chemical composition during inward pebbles drift and the subsequent elemental composition of the planet formed. In this scenario, the grains migrate and carry the frozen CO or CO$_2$ on the surface to their respective snowline, increasing locally the gas phase abundances \citep[e.g.][]{RBooth+2017}. Volatile trapping should reduce the chemical enhancement at snowline locations while the drifting pebble surfaces retain more CO and CO$_2$ ice than predicted. Superstellar C/O ratios can still be explained by the accretion of gases, although volatile trapping does reduce the amount of available gaseous material. 

%--------------------------------------------------------------------
\section{Conclusions}

In this study, we demonstrate that volatile trapping in bulk ice strongly affects C/O ratios in protoplanetary disks. Strong variations in the C/O ratios seen in models without trapping have been reduced. Solid elemental ratios become constant from outside the CO snowline to inside the H$_{2}$O snowline. While gas phase C/O ratios are still enhanced, trapping reduces the amount of carbon- and oxygen-bearing molecules available in the gas. A similar effect is expected for volatile molecules that set the budget of other elements, such as nitrogen and sulfur. The homogenisation of the C/O ratios ensures that planets can acquire a stellar C/O ratio outside and inside the CO snowline. Observations of C/O ratios in exoplanet atmospheres are therefore difficult to link to a specific formation location in a protoplanetary disk. 

\begin{acknowledgements}
    The authors thank E.G. Bøgelund, S.F. Wampfler, J. Drążkowska, D. Semenov, H. Cuppen and M.N. Drozdovskaya for useful discussions, and A. Cridland for feedback on the manuscript. N.F.W.L. and K.A.K. acknowledge support from the Swiss National Science Foundation (SNSF) Ambizione grant 193453 and NCCR PlanetS. S.G. acknowledge support from the Independent Research Fund Denmark (grant No. 0135-00123B).
\end{acknowledgements}

\bibliographystyle{aa}
\bibliography{bib}

%--------------------------------------------------------------------
\begin{appendix}

\section{Trapping fractions}
\label{ap:trapping}

Table \ref{tab:frac} shows the trapped fractions of volatiles obtained from various studies. Trapping fractions are usually determined in the following way. In a laboratory setup, an ice film consisting of two or more components is prepared by dosing gas mixture on an ultra-cold (5 -- 80~K) surface. Next, the ice is linearly heated in a process called Temperature Programmed Desorption (TPD) with heating rates that are generally a few K~min$^{-1}$. A mass spectrometer records the molecules desorbing from the ice film at their characteristic mass-over-charge ($m/z$), for example, $m/z$=18 for H$_{2}$O$^{+}$, $m/z$=28 for CO$^{+}$, and $m/z$=44 for CO$_{2}^{+}$, see Fig. \ref{fig:tpd}. From the TPD trace, desorption and trapping fractions are determined by integrating the signal over a specified temperature regime that corresponds to a certain desorption event. For example, \citet{kipfer2024sublimation} determines CO desorption fractions by integrating between 15--60~K for pure CO, 60--115~K for CO co-desorption with CO$_{2}$, and 115--200~K for CO release associated with water (this combines the so-called volcano desorption event and the water co-desorption into one). Between different studies, integration ranges can differ slightly, but this does not significantly affect determined fractions due to the intensity of desorption events (note the logarithmic scale in Fig. \ref{fig:tpd}). 

Various studies provide relevant TPD data, but without determining desorption or trapping fractions. In these cases, fractions can be retrieved by analysing digitised TPD traces, see for example \citet{Rubin_2023}, where fractional desorption is determined for TPD data from \citet{kouchi1995cosmoglaciology} and \citet{Gudipati_2023}. For this study, we additionally analyse TPD data from \citet{collings2004} and \citet{Martin_Domenech_2014}. Integration range are set from $T_{\rm start}$ -- 60~K for pure CO desorption, 60 -- 130~K for pure CO$_{2}$ desorption and CO co-desorption, and 130 -- 200~K for CO and CO$_{2}$ release associated with H$_{2}$O sublimation, where $T_{\rm start}$ is the lowest temperature presented in the TPD trace. 

Some nuances are important to be aware of when interpreting desorption data of ice films consisting of three or more components. The desorption behaviour of binary ice mixtures is fairly easy to understand. Usually, a volatile species (e.g. CO) is embedded in a more abundant and less volatile host medium (e.g. CO$_{2}$). In these cases, the release of the volatile occurs at its pure-phase sublimation temperature and when the host medium sublimates \citep[for example, see][for experiments of CO trapping in CO$_{2}$]{simon2019entrapment}. Trends in trapping fractions are easy to assess based on mixing ratios and ice film thicknesses. However, when a third component is added such as H$_{2}$O, the trapping behaviour of CO in CO$_{2}$ is changed, because CO$_{2}$ itself interacts with and is trapped in H$_{2}$O \citep[see e.g.][]{kipfer2024sublimation}. This complicates plotting trends of the CO co-release with CO$_{2}$, because the pure-phase CO$_{2}$ thickness and mixing ratios are difficult to determine and affected by the additional components in the ice. Therefore, the total ice thickness and mixing ratios are used to plot trends in trapping behaviour in ice mixtures consisting of three or more components, but this is not an accurate representation of the ice composition. 

\begin{table*}
\caption{Volatile trapping experimental results}             
\label{tab:frac}      
\centering                          
\begin{tabular}{c c c c l}        
\hline\hline                 
Molecule & $F_{\rm H_{2}O}$(X) & Thickness (ML) & H$_{2}$O:X & Reference \\
\hline    
CO & [0.41 -- 0.64] & 1000 & [21 -- 125] & \citep{kipfer2024sublimation} \\
CO & 0.23 & 1000 & 6 & \citet{Gudipati_2023}, \citet{Rubin_2023} \\ 
CO & [0.27 -- 0.61] & [14 -- 69] & [3 -- 10] & \citet{Simon_2023} \\   
CO & [0.36 -- 0.61] &  [309 -- 1396] &  [6.7--11.5] &\citet{Martin_Domenech_2014}$^{a}$ \\
CO & [0.04 -- 0.96] & [14 -- 30] & [1 -- 20] & \citet{fayolle2011laboratory} \\
CO & 0.21 & 105 & 20 & \citet{collings2004}$^{a}$ \\
CO & 0.07 & $\geq$1000 & 4.3 & \citet{kouchi1995cosmoglaciology}, \citet{Rubin_2023} \\
\hline
CO$_{2}$ & 0.55 & 1000 & 3 & \citet{Gudipati_2023}, \citet{Rubin_2023} \\
CO$_{2}$ & [0.76 -- 0.94] & [750 -- 1396] & [5 -- 10]  & \citet{Martin_Domenech_2014}$^{a}$ \\
CO$_{2}$ & [0.32 -- 0.92] & [14 -- 30] & [3 -- 20] & \citet{fayolle2011laboratory} \\
CO$_{2}$ & 0.34  & 105 & 20 & \citet{collings2004}$^{a}$ \\
CO$_{2}$ & 0.20 & $\geq$1000 & 6.5 & \citet{kouchi1995cosmoglaciology}, \citet{Rubin_2023} \\
\hline
\hline
Molecule & $F_{\rm CO_{2}}$(CO) & Thickness (ML) & CO$_{2}$:CO & Reference \\
\hline
CO & [0.11 -- 0.34] & 1000 & [2 -- 28] & \citet{kipfer2024sublimation}\\
CO & 0.07 & $\geq$1000 & 0.67 & \citet{Gudipati_2023}, \citet{Rubin_2023} \\ 
CO & [0.33 -- 0.67] & [13 -- 72] & [3 -- 13] & \citet{Simon_2023} \\
CO & [0.14 -- 0.60] & [6 -- 53] & [1 -- 9] & \citet{simon2019entrapment} \\
CO & 0.17 & 1000 & 2 & \citet{kouchi1995cosmoglaciology}, \citet{Rubin_2023} \\
\hline                                   %inserts single line
\end{tabular}
\tablefoot{$F_{\rm H_{2}O}$(x) indicates the fraction of CO or CO$_{2}$ trapped in water. $F_{\rm CO_{2}}$(CO) indicates the fraction of CO trapped in CO$_{2}$. The thickness is indicated in monolayers (ML), where 1 ML = 10$^{15}$ molecules cm$^{-2}$. $^{a}$Trapping fractions determined in this work.}
\end{table*}

\newpage

\section{Additional plots}
\label{ap:plots}

\begin{figure}[!h]
\centering
\includegraphics[width=\hsize]{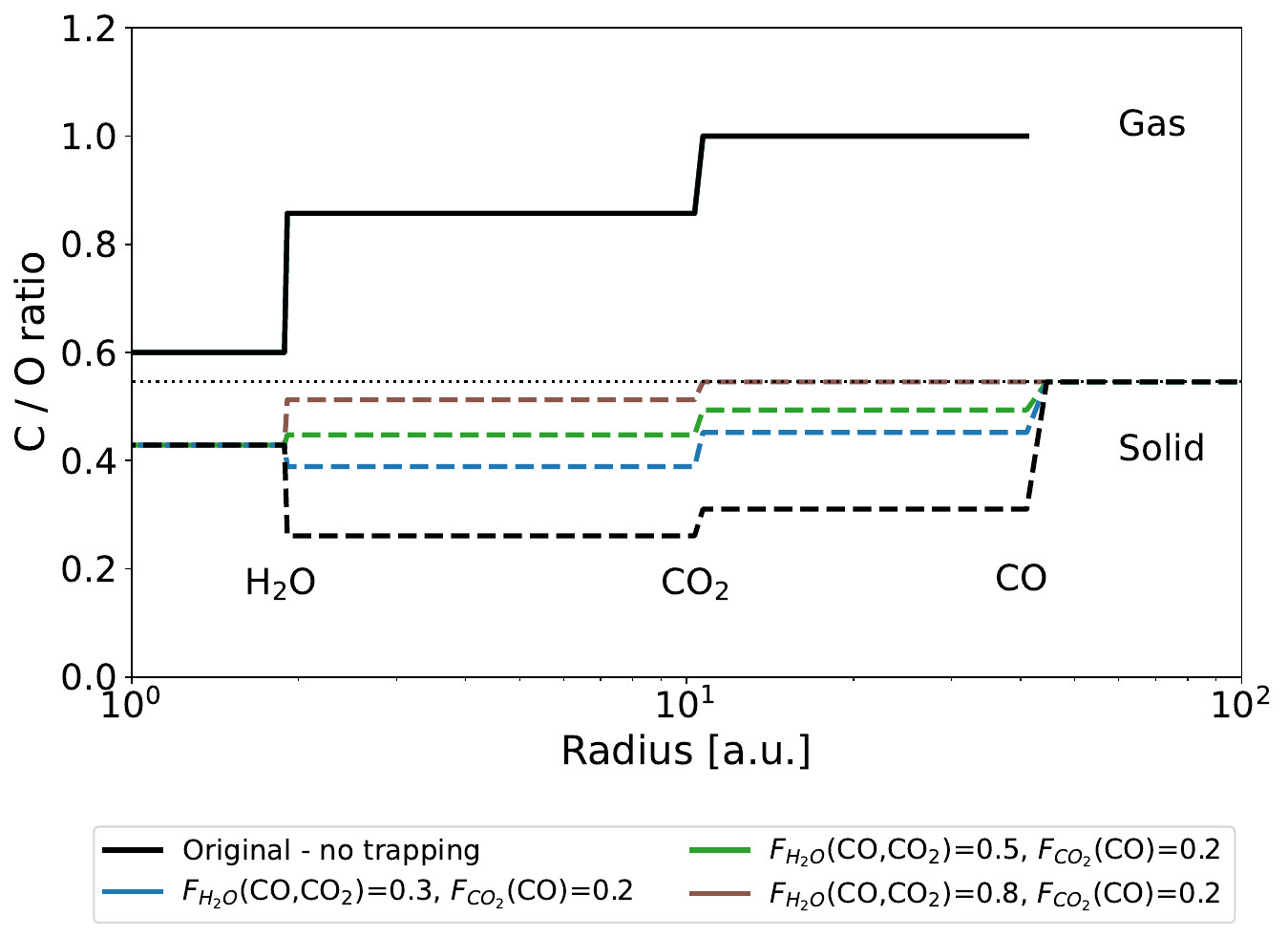}
  \caption{Same as Figure \ref{fig:1}, but including CO trapping in CO$_{2}$. We note that for the model $F_{\rm H_{2}O}$(CO,CO$_{2}$) = 0.8, $F_{CO_{2}}$(CO) = 0.2 there is no gas phase C/O ratios outside the CO$_{2}$ snowline because CO does not release before the CO$_{2}$ co-desorption event.
          }
     \label{fig:3}
\end{figure}

\begin{figure}[!h]
\centering
\includegraphics[width=\hsize]{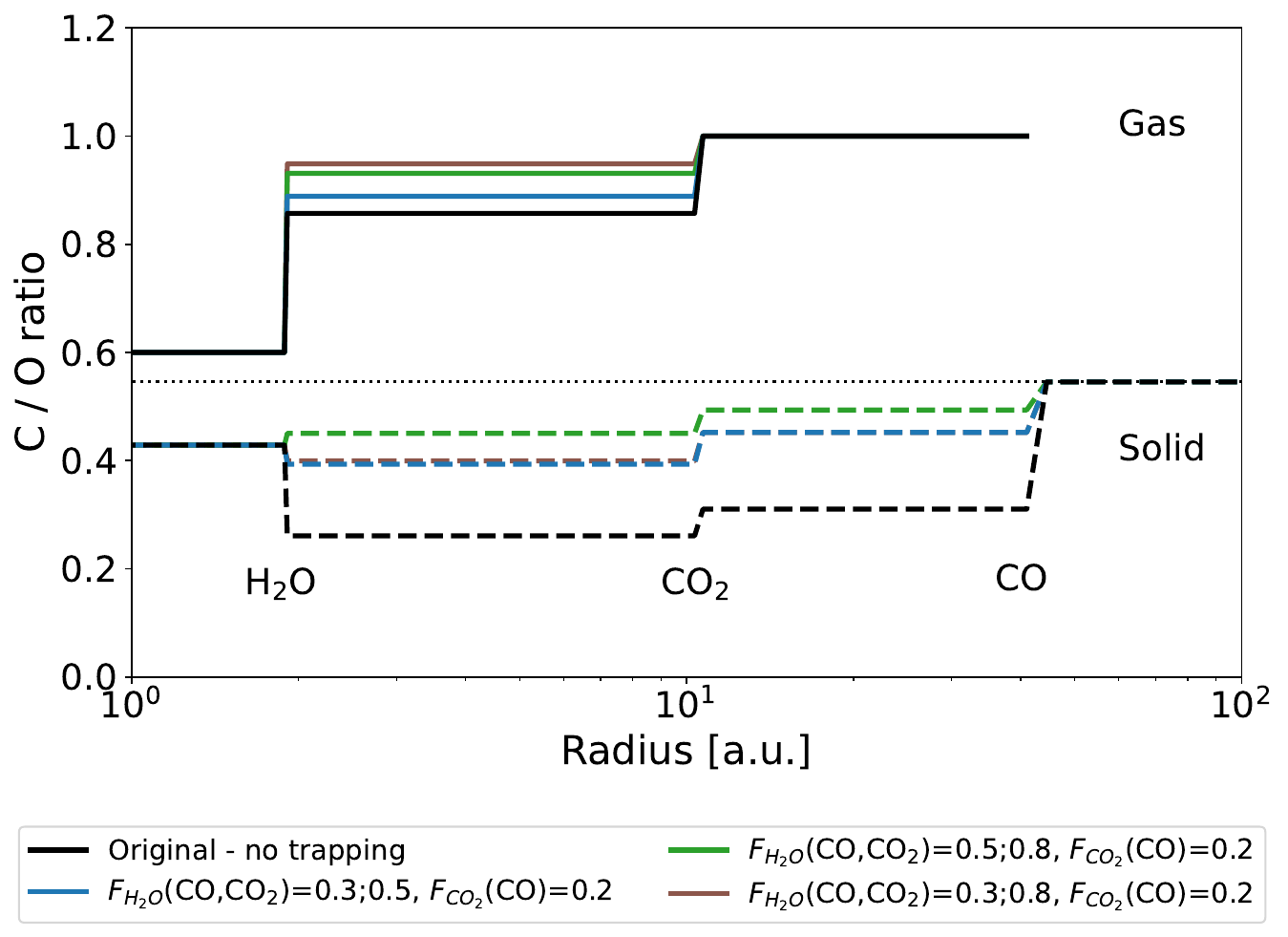}
  \caption{Same as Figure \ref{fig:1}, but different trapping ratios of CO and CO$_{2}$ in H$_{2}$O and CO trapping in CO$_{2}$.
          }
     \label{fig:4}
\end{figure}

Additional model results where not only volatile trapping in H$_{2}$O is included, but also CO trapping in CO$_{2}$ are shown in Figs. \ref{fig:3} and \ref{fig:4}. A reduction in C/O variations is again observed, but the release of a fraction of the CO with CO$_{2}$ causes a more pronounced drop in solid-state C/O ratio around the CO$_{2}$ snowline. 

\end{appendix}

\end{document}